%Paper: cond-mat/9301004
%From: jysoo@hlrserv.hlrz.kfa-juelich.de (Jysoo Lee)
%Date: Wed, 6 Jan 93 08:27:26 +0100

\magnification=1200
\baselineskip=20pt
\normallineskip=10pt
\overfullrule=0pt
\footline={\sevenbf LZ4741 --- Jan 5 '93 \hss PAGE \folio}

\null
\hfill HLRZ preprint 101/92
\vfill
\centerline {\bf Heap Formation in Granular Media}
\bigskip
\bigskip

\centerline {Jysoo Lee}
\bigskip

\centerline {HLRZ-KFA J\"{u}lich, Postfach 1913, W-5170 J\"{u}lich, Germany}
\bigskip
\vfill

\centerline{\bf Abstract}
\bigskip

Using molecular dynamics (MD) simulations, we find the formation of
heaps in a system of granular particles contained in a box with
oscillating bottom and fixed sidewalls. The simulation includes the
effect of static friction, which is found to be crucial in maintaining
a stable heap. We also find another mechanism for heap formation in
systems under constant vertical shear. In both systems, heaps are
formed due to a net downward shear by the sidewalls. We discuss the
origin of net downward shear for the vibration induced heap.
\bigskip


\vfill
\eject
\null

Systems of granular particles (e.g. sand) exhibit many interesting
phenomena, such as segregation under vibration or shear, density waves
in the outflow through hoppers, and probably most strikingly, the
formation of heap and convection cell under vibration.$^{1-4}$ It has
been known for more than one hundred years that granular particles on
the top of a vibrating surface will form convection cells and
heaps.$^{5}$ However, even with many recent studies on the
subject,$^{6-11}$ the exact mechanism for the heap formation is not
established.
\medskip

Recently, two experimental groups, Evesque {\it et al}$^{6}$ and
Laroche {\it et al},$^{7}$ studied behaviors of granular particles
contained in a box, while the whole box is being vertically vibrated.
They confirm the formation of convection cell and heap. On the other
hand, Zik {\it et al} find convection but no heap.$^{8}$ When viewed
from above, these boxes are essentially squares, making the system
fundamentally $3$-dimensional. On the other hand, there are some
studies in $2$ dimension (i.e. a line when viewed from above) with
fruitful results.  Using molecular dynamics (MD) simulations of
granular particles, Taguchi$^{9}$ and Gallas {\it et al}$^{10}$ found
convection cells under vibration in $2$ dimension. Furthermore, they
established the fact that the sidewalls are {\it inducing} the
convection. However, the exact mechanism of how the convection is
induced by the wall is still not firmly established. Also, they did
not find any formation of heap. Another breakthrough is the
experimental discovery of heap formation in $2$ dimension by Clement
{\it et al}.$^{11}$ Using monodisperse particles, they found that (1)
the static friction coefficient must be relatively large in order to
induce convection and heap, and (2) the heap is formed as particles
are being pushed upward by the sidewalls (the wall induces convection)
along the surface, while there is no significant motion in the bulk.
The lack of motion in the bulk is probably the consequence of
hexagonal packing due to monodispersity, and not an essential part of
the heap formation.
\medskip

The very reason why granular particles can form a stable pile is
static friction. More precisely, the contact between particles must be
able to withstand a finite amount of shear force in order to maintain
a pile. We implement static friction in MD simulations of granular
material using the scheme of Cundall and Strack.$^{12}$ In this
scheme, one has to apply a {\it finite} force in order to break a
contact between particles. Using the implementations, we study heap
formations in $2$ dimension. First, we present heap formations due to
shear (``the shear induced heaping''), which are intimately connected
to ``the vibration induced heaping.'' We study the situation that
sidewalls are moving vertically in opposite directions with constant
velocity, thereby creating asymmetrical shear in the cell. Here, the
bottom plate is not moving. We find the formation of convection {\it
and heap}. In these simulations, the walls are dragging nearby
particles, which causes a net flux of particles. This flux is inducing
convection in the cell, and the convection builds a heap, which is
stable due to the presence of static friction. We also study the
parameter dependence of the formation, and find the two static
friction coefficients, one between the wall and a particle and the
other between particles, are the most important. We next study the
case that {\it both} walls are moving down with constant velocity,
which causes symmetric shear. We also find a convection cell and
heaping, whose formations are essentially the same as the asymmetric
case. Finally, we study the case of vibration induced heaping. We
first fix the sidewalls and vibrate the bottom plate. We find heap
formation and convection for a range of amplitude and frequency. Based
on several measurements, we propose the following mechanism for the
formation.  The bottom plate is moving up or down during one half of a
cycle. The density of particles are found to be smaller during the
downward phase, which cause the shear force by the walls to be larger
in absolute magnitude during the upward phase.  Over one cycle, the
net shear force applied by the wall is downward, which cause net
downward flux of particles near the walls. Therefore, the situation is
very similar to the case of the symmetric shear. We also study the
case of vibrating the sidewalls as well as the bottom plate, and find
convection but no heap. We discuss possible explanation.
\bigskip

The force between two particles $i$ and $j$, in contact with each
other, is the following. Let the coordinate of the center of particle
$i$ ($j$) to be $\vec{R}_i$ ($\vec{R}_j$), and $\vec{r} = \vec{R}_i -
\vec{R}_j$.  In two dimension, we use a new coordinate system defined
by the two vectors $\hat{n}$ (normal) and $\hat{s}$ (shear). Here,
$\hat{n} = \vec{r} / {\vert \vec{r} \vert}$, and $\hat{s}$ is defined
as rotating $\hat{n}$ clockwise by $\pi/2$.  The normal component
$F_{j \to i}^{n}$ of the force acting on particle $i$ by $j$ is
$$
F_{j \to i}^{n} = k_n (a_i + a_j - \vert \vec{r}
\vert)^{3/2} - \gamma_n m_e (\vec{v} \cdot \vec{n}),
\eqno (1a)
$$
\noindent
where $a_i$ ($a_j$) is the radius of particle $i$ ($j$), and $\vec{v}
= d\vec{r}/dt.$ The first term is the Hertzian elastic force, where
$k_n$ is the elastic constant of the material. And, the constant
$\gamma_n$ of the second term is the friction coefficient of a
velocity dependent damping term, $m_e$ is the effective mass, $m_i
m_j/(m_i + m_j).$ The shear component $F_{j \to i}^{s}$ is given by
$$
F_{j \to i}^{s} = - \gamma_s m_e (\vec {v} \cdot \vec {s}) - {\rm
sign} (\delta s) ~ {\rm min}(k_s \vert \delta s \vert, \mu \vert F_{j
\to i}^n \vert),
\eqno (1b)
$$
\noindent
where the first term is a velocity dependent damping term similar to
that of Eq.~(1a). The second term is to simulate static friction,
which requires a {\it finite} amount of force ($\mu F_{j \to i}^{n}$)
to break a contact.$^{12}$ Here, $\mu$ is the friction coefficient,
$\delta s$ the {\it total} shear displacement during a contact, and
$k_s$ the elastic constant of a virtual spring. There are several
studies on granular systems using the above interactions.$^{13}$
However, only a few of them,$^{12,14,15}$ include static friction. A
particle can also interact with a wall. The force on particle $i$, in
contact with a wall, is given by Eqs.~(1) with $a_j =
\infty$ and $m_e = m_i$.  A wall is assumed to be rigid, i.e. it is
not affected by collisions with particles. Also, the system is under a
gravitational field $\vec{g}$. We do not include the rotation of the
particles in present simulation. A detailed explanation of the
interaction is given elsewhere.$^{15}$
\medskip

We first consider the situation that systems of granular particles are
under constant vertical shear. Consider a box of width $W$ and height
$H$. We insert particles at randomly chosen positions inside the box,
and calculate the trajectories of the particles by a fifth order
predictor-corrector method. The particles fall by gravity, lose their
energy through collisions, and fill the box without any significant
motion.  The parameters we use for the interaction between the
particles are $k_n = 1.0 \times 10^6, k_s = 1.0 \times 10^4$,
$\gamma_n = 1.0 \times 10^3, \gamma_s = 0$ and $\mu_{pp} = 0.2$. For
the interaction between the particle and the wall, we use $k_n = 2.0
\times 10^6$, $k_s = 1.0 \times 10^4$, $\gamma_n = 5.0 \times 10^2$,
$\gamma_s = 0$. The friction coefficient at the sidewall and bottom
plate are $\mu_{pw} = 5.0$ and $0.2$, respectively. The time step is
chosen to be $5 \times 10^{-5}$, and gravity $g$ is $980$. In this
letter, CGS units are implied. In order to avoid the hexagonal packing
formed by particles of the same size, we choose the radius from a
gaussian distribution with average $0.1$ and width $0.02$. The density
of the particles is chosen to be $0.5$.  Later, we also study the
system of monodisperse particles. We then apply a vertical shear by
pulling the right (left) wall with constant velocity $v_{s} = 0.2$
($-0.2$). In Fig.~1(a), we show the system after $80 000$ iterations
of the vertical shear. The slope of the surface of the pile increases,
and fluctuates around a non-zero value. The mechanism to generate the
heap is rather simple.  Since one pulls the sidewalls with constant
velocity, the walls exert shear forces to nearby particles. If the
force at the wall is sufficiently high, it will induce flow of
particles in the vertical direction. The upward (downward) flow of
particles near the right (left) wall, combined with static friction,
results in the formation of the heaps.
\medskip

We study the effect of parameters on the formation of the heaps. There
are quite a few parameters in the system. However, most parameters,
while their values are chosen within reasonable ranges, do not affect
the behavior of the system. The key parameters are the two friction
coefficients $\mu_{pw}$ and $\mu_{pp}$, and the shear velocity of the
sidewalls $v_{s}$.  First, we study the effect of $\mu_{pw}$. We fix
$\mu_{pp} = 0.2$, $v_{s} = 0.2$, and the friction coefficient of the
bottom plate to be zero. In Fig.~1(b), we show the average angle of
the pile $\langle \theta \rangle$ for different values of $\mu_{pw}$
with $W=3$ and the number of particles $n=150$. Here, averages are
taken over time (excluding the transient) and several different runs,
where each angle is averaged over approximately $5000$ points. For
small $\mu_{pw}$ ($0.5$ or $1.0$), the particles do not move
significantly during the whole simulations, which results in a zero
angle. In order to have convective motion and heaping, $\mu_{pw}$
should be larger than certain threshold $\mu_{pw}^{c}$. The existence
of a finite threshold $\mu_{pw}^{c}$ can be understood as follows. In
order to lift particles near the right wall, the shear force by the
right wall should be larger than the sum of the gravitational force
and the friction between particles.  Since the sum is finite, one
needs finite $\mu_{pw}$ in order to maintain the convection. It is
still an open question whether the transition is the first or the
second order, i.e., whether there exists a sudden jump of $\langle
\theta \rangle$.
\medskip

We now fix $\mu_{pw} = 0.2,$ $v_{s} = 1.0$, and study the effect of
$\mu_{pp}$. We calculate $\langle \theta \rangle$ for several values
of $\mu_{pp}$, where the averages are taken over approximately $1000$
points.  Here, $W=3$ and $n=150$. The angle $\langle \theta \rangle$
becomes larger for larger values of $\mu_{pp}$, which may results from
the fact that the angle of repose is an increasing function of
$\mu_{pp}$.$^{15}$ We then study the effect of $v_{s}$ by fixing
$\mu_{pw} = 5.0,~\mu_{pp} = 0.2$, and change $v_{s}$. We measure
$\langle \theta \rangle$ for several values of $v_{s}$ between $0.1$
and $10.0$ with $W=3,~n=150$.  The measured angle is quite insensitive
to $v_{s}$. For example, the angle is $25.8$ for $v_{s} = 0.1$ and
$23.6$ for $v_{s} = 10.0$. When $v_{s}$ is increased, the pile tries
to increase the slope due to larger current of particles. On the other
hand, increased motion of particles decreases the stablizing effect of
static friction. These two effects seem to cancel each other resulting
to the insensitive dependence.  Finally, we study the same system
using monodisperse particles. The system just before we apply the
shear is a hexagonal packing of particles with few defects. When the
shear of $v_{s} = 0.2$ is applied, the hexagonal packing becomes
unstable, and the system starts to form a square packing.  When the
formation of the square packing is completed, the particles near the
wall can withstand the applied shear with no motion. The square
packing, which is stable for small values of shear, becomes unstable
as $v_{s}$ is increased.
\medskip

So far, we have studied the formation of heaps by an asymmetric shear,
i.e., the sidewalls are moving in the {\it opposite} direction. We now
consider the case of a symmetric shear, where both sidewalls are
moving in the {\it same} direction. In Fig.~1(c), we show the system
after $50 000$ iterations. Here, we use $\mu_{pw} = 5.0,~\mu_{pp} =
0.2$, and both walls are moving down with constant velocity $v_{s} =
-1.0$. The mechanism of generating the symmetric heap shown in the
figure is essentially the same as that of the asymmetric heap. The
shear force induces downward flow of particles near the sidewalls. The
flow merges together around the center of the cell, and rises to the
top of the pile.
\medskip

We want to argue that the above ``shear induced heaping'' is related
to the ``vibration induced heaping.'' In fact, the above shear
geometries are chosen to demonstrate more clearly their similarity. We
now study the vibration induced heaping, and discuss its relation with
the shear induced case. We first {\it fix} both sidewalls of a box and
vibrate the bottom plate with amplitude $A$ and frequency $f$. In
Fig.~2, we show the system after $16$ cycles as well as the
displacements of the particles over $15$ cycles. For this simulation,
we take $W = 10$, $n = 800$, $A = 0.190$ and $f = 20~Hz$. The
parameters of the interaction of the particles and the sidewall are
$\mu = 3.0,$ $k_s = 1.0 \times 10^6$. For the interaction between the
particles, we use $\mu = 0.5$ and $k_s = 5.0 \times 10^4$.  All the
other parameters are kept the same as before. In the figure, one can
clearly see a heap and associated convection.
\medskip

We now discuss the mechanism for the formation of the heap. In Fig.~2,
we show the average number of particles $c(\phi )$ in contact with one
particle for various phases $\phi$ during one cycle.  Here, $\phi $ is
in the unit of $2 \pi$, and the averages are taken over $20$ cycles.
The numbers $c(\phi )$ are smaller during the downward phase ($0.25 <
\phi < 0.75$) than the upward phase. One of the consequences of this
``up/down symmetry breaking'' is the shear forces of the sidewalls are
also asymmetric. In Fig.~2, we show the total shear force $f_{s}(\phi
)$, which the right wall applies to the particles, for several values
of $\phi$. The sign of $f_{s}$ is roughly opposite to that of the
velocity of the bottom. The absolute magnitude of $f_{s}$ is larger
for the upward phase, and because the particles are more densely
packed, the wall can exert a larger force.  Since the shear force is
essentially a drag force for the particles, we expect particles move
faster vertically during the downward phase, where the shear force is
smaller. Therefore, there is net downward flux of particles near the
sidewalls, which results in a convective motion and heaping.  In
summary, the convection and heaping is due to the net current along
the sidewalls, which is caused by the net downward shear, which again
is a result of ``up/down symmetry breaking'' of the particle density.
\medskip

We consider a similar case when we vibrate the sidewalls as well as
the the bottom plate. We find that the above mechanism still
holds---there is net downward shear and flux of particles near the
wall, and convection cells. However, we {\it do not} find a heap. To
understand this, we note that the vibration, which is the source of
convection and heaping, also tends to make an existing heap unstable.
The shear force, which is the driving force of the convection, is
proportional to the relative shear displacement between the walls and
particles due to the present implementation scheme of static friction.
With fixed walls, we can obtain rather large net shear force for small
$A$, even when the acceleration of the bottom is smaller than $g$.
With moving walls, however, we need the acceleration to be at least as
large as $g$ to have the relative displacement and net shear. For
large $A$, the vibration induces motion of particles along the top
surface, which dominates the current due to the convective motion.
This destroys any existing heap. This observation does not necessarily
imply that there is no heap formation with moving walls in
$2$-dimension. It has been known that there are some range of $A$
(``windows'') for the formation of a heap.$^{16}$ The width of the
window is very likely to be dependent on the parameters of the
material. The window may be very narrow for the range of parameters we
are studying. A similar parameter problem could also explain why no
heap was found in the experiment by Zik {\it et al}.$^{8}$. There is
limitation on the range of parameters we can simulate. For example,
$k_{s}$ is very large (ideally, infinity) for real material, and
larger $k_{s}$ is more effective in creating convection. However, we
can not simulate values of $k_{s} > 1.0 \times 10^{6}$
without significantly decreasing the time step. The situation is
entirely similar with $\gamma_{n}$---we can not increases $\gamma_{n}$
beyond $1 \times 10^{3}$. As a result, it is possible that the heap
can be found in these parts of the parameter space.
\medskip

It has been observed previously that walls are responsible for the
convection and/or the formation of heaps, and there have been
conflicting arguments on the way how the walls {\it induce} the
convection.$^{9-10}$ We presented here an argument based on measuring
properties of the system. Our argument is similar to that of Gallas
{\it et al}$^{10}$ in the sense that both are based on the shear force
that the walls are exerting on the particles. However, the two
theories have different mechanisms for the generation of the net shear
force.
\medskip

In conclusion, we find heap formations in two types of systems---one
with constant vertical shear, the other a with vibrating bottom and
fixed walls. Heaps in both systems are caused by a net downward shear.
Both of the systems are not studied experimentally, and there are many
interesting quantities to measure. In the vibration induced heaping,
it would be nice to check for the existence of a net shear by
measuring the shear stress of the walls. It would be important to
study the parameter dependence of the angle of repose. In the shear
induced case, the further understanding of a parameter dependences of
$\langle \theta \rangle$ (especially, $\mu_{pw}$) is necessary
experimentally as well as theoretically.  Unfortunately, heap
formation in $3$ dimension can not be explained by this mechanism,
since it is known that heaps can be formed without a boundary in
$3$-d.  The mechanism for $3$-d heap formation still remains to be
understood.
\bigskip

I thank Hans Herrmann and Michael Leibig for many useful discussions.
\bigskip

\vfill
\eject
\null
{\bf References}
\bigskip

\item{1.} S. B. Savage, Adv. Appl. Mech. {\bf 24}, 289 (1984); S. B.
Savage, {\it Disorder and Granular Media} ed. D.  Bideau,
North-Holland, Amsterdam (1992).
\medskip

\item{2.} C. S. Campbell, Annu. Rev. Fluid Mech. {\bf 22}, 57 (1990).
\medskip

\item{3.} H. M. Jaeger and S. R. Nagel, Science {\bf 255}, 1523
(1992).
\medskip

\item{4.} A. Mehta, Physica A {\bf 186}, 121 (1992).
\medskip

\item{5.} M. Faraday, Phil. Trans. R. Soc. London {\bf 52}, 299
(1831).
\medskip

\item{6.} P. Evesque and J. Rajchenbach, Phys. Rev. Lett. {\bf62},
44 (1989).
\medskip

\item{7.} C. Laroche, S. Douady, S. Fauve, J. de Physique {\bf 50},
699 (1989).
\medskip

\item{8.} O. Zik and J. Stavans, Europhys. Lett. {\bf 16}, 255 (1991).
\medskip

\item{9.} Y.-h. Taguchi, Phys. Rev. Lett. {\bf 69}, 1371 (1992).
\medskip

\item{10.} J. A. C. Gallas, H. J. Herrmann and S. Soko\l owski, Phys.
Rev. Lett. {\bf 69}, 1375 (1992).
\medskip

\item{11.} E. Clement, J. Duran and J. Rajchenbach, Phys. Rev. Lett.
{\bf 69}, 1189 (1992).
\medskip

\item{12.} P. A. Cundall and O. D. L. Strack, G\'{e}otechnique {\bf
29}, 47 (1979).
\medskip

\item{13.} For example, P. K. Haff and B. T. Werner, Powder Technol.
{\bf 48}, 239 (1986); P. A. Thompson and G. S. Grest, Phys. Rev. Lett.
{\bf 67}, 1751 (1991); G. Ristow, J. Physique I, {\bf 2}, 649 (1992).
\medskip

\item{14.} Y. M. Bashir and J. D. Goddard,  J. Rheol. {\bf 35}, 849
(1991).
\medskip

\item{15.} J. Lee and H. J. Herrmann, HLRZ preprint 44/92.
\medskip

\item{16.} S. Douady, S. Fauve and C. Laroche, Europhys. Lett. {\bf
8}, 621 (1989).
\vfill
\eject

\null
{\bf Figure Captions}
\bigskip

\item {Fig.~1:} Shear induced heap formations: (a) Configuration after
$80 000$ iterations of asymmetric shear where the right (left) wall is
moving up (down) with constant velocity $v_{s} = 0.2$. (b) The
dependence of the angle of repose $\langle \theta \rangle$ on the
friction coefficient of the wall $\mu_{pw}$. (c) Configuration after
$50 000$ iterations of symmetric shear, where both walls are moving
down with constant velocity $v_{s} = -1.0$.
\medskip

\item {Fig.~2:} Vibration induced heap formation: Configuration
after $16$ cycles of vibration with vibrating bottom plate and fixed
sidewalls. Displacements of particles over $15$ cycles are also shown.
In the insets, we also show the average number of contact $c(\phi )$
and the shear force $f_{s}(\phi )$ for the different phases $\phi$.
\medskip

\vfill
\eject
\bye